\documentstyle[twocolumn,aps,prl,epsf,psfig,bbm]{revtex}
\begin{document}
\def\r{{\bf{r}}}
\def\i{{\bf{i}}}
\def\j{{\bf{j}}}
\def\m{{\bf{m}}}
\def\k{{\bf{k}}}
\def\kt{\tilde{\k}}
\def\K{{\bf{K}}}
\def\P{{\bf{P}}}
\def\q{{\bf{q}}}
\def\Q{{\bf{Q}}}
\def\p{{\bf{p}}}
\def\x{{\bf{x}}}
\def\X{{\bf{X}}}
\def\Y{{\bf{Y}}}
\def\F{{\bf{F}}}
\def\G{{\bf{G}}}
\def\M{{\bf{M}}}
\def\V{\cal V}
\def\tchi{\tilde{\chi}}
\def\tk{\tilde{\bf{k}}}
\def\tK{\tilde{\bf{K}}}
\def\tq{\tilde{\bf{q}}}
\def\tQ{\tilde{\bf{Q}}}
\def\si{\sigma}
\def\ep{\epsilon}
\def\ek{\epsilon_{\bf k}}
\def\eK{\bar{\epsilon}_{\bf K}}
\def\al{\alpha}
\def\bet{\beta}
\def\ep{\epsilon}
\def\up{\uparrow}
\def\de{\delta}
\def\De{\Delta}
\def\up{\uparrow}
\def\dwn{\downarrow}
\def\ksi{\xi}
\def\etha{\eta}
\def\product{\prod}
\def\goto{\rightarrow}
\def\switch{\leftrightarrow}
\def\be{\begin{equation}}
\def\ee{\end{equation}}
\def\bea{\begin{eqnarray}}
\def\eea{\end{eqnarray}}

\title{d-wave Superconductivity in the Hubbard Model}
\author{Th.\ Maier$^1$, M.\ Jarrell$^2$, Th.\ Pruschke$^1$, J.\ Keller$^1$}
\address{
$^1$Institut f\"ur Theoretische Physik, Universit\"at Regensburg,
93040 Regensburg\\
$^2$Department of Physics, University of Cincinnati, Cincinnati, OH 45221-0011}
\date{\today }
\maketitle

\begin{abstract}
The superconducting instabilities of the doped repulsive 2D Hubbard 
model are studied in the intermediate to strong coupling regime
with help of the Dynamical Cluster Approximation (DCA). To solve the
effective cluster problem we employ an extended Non
Crossing Approximation (NCA), which allows for a transition to the
broken symmetry state. At sufficiently low temperatures we find stable 
d-wave solutions with off-diagonal long range order. The maximal $T_c\approx 150K$
occurs for a doping $\delta\approx 20\%$ and the doping dependence of
the transition temperatures  agrees well with the
generic high-$T_c$ phase diagram.
\end{abstract}


\paragraph*{Introduction} 
The discovery of high-$T_c$ superconductors has stimulated strong
experimental and theoretical interest in the field of strongly
correlated electron systems. After a decade of intensive studies we
are still far from a complete understanding of the rich physics
observed in high-$T_c$ cuprates \cite{maple1}. Angle resolved photoemission experiments on doped materials show a $d$-wave
anisotropy of the pseudogap in the superconducting state \cite{ding}. In
underdoped materials even in the normal state this pseudogap persists \cite{ding,ronning},
which is believed to cause the unusual non-Fermi-liquid
behavior in the normal state. This emphasizes the importance of
achieving a better understanding of the superconducting phase,
i.e. the physical origin of the pairing mechanism, the nature of the pairing
state and the character of low energy excitations. 

On a phenomenological basis the $d$-wave
normal state pseudogap as well as the transition to a superconducting
state with a $d$-wave order parameter has been described within
theories where short-ranged antiferromagnetic spin fluctuations mediate
pairing in the cuprates \cite{scalapino1,Monthoux,timusk}.       

On a microscopic level it is believed that the Hubbard model or
closely related models like the t-J model should capture the essential
physics of the high-$T_c$ cuprates \cite{Anderson}. However, despite
years of intensive studies, these models remain unsolved except in one 
or infinite dimensions.

Finite size QMC calculations for the doped 2D Hubbard model in the 
intermediate coupling regime with Coulomb repulsion $U$ less than
or equal to the bandwidth $W$, support the idea of a spin fluctuation 
driven interaction mediating $d$-wave superconductivity \cite{scalapino1}. 
But the fermion sign problem limits these calculations to temperatures 
too 
high to observe a
possible Kosterlitz-Thouless transition for the 2D system \cite{scalapino1}. 
Another problem encountered in QMC calculations is their finite size 
character, which makes statements for the thermodynamic limit dependent on 
a scaling ansatz.

These limitations do not apply to approximate many particle methods like 
the Fluctuation Exchange Approximation (FLEX) \cite{bickers,Moriya}. Results 
of FLEX calculations for the Hubbard model are in agreement with QMC 
results, i.e. they show evidence for a 
superconducting state with $d$-wave order parameter at moderate doping for 
sufficiently low temperatures\cite{bickers,Moriya}. But the FLEX method as 
an approximation based on a perturbative expansion in $U$ breaks down in 
the strong coupling regime $U>W$, where $W$ is the bare bandwidth. On the other hand it is believed that a proper 
description of the high-$T_c$ cuprates in terms of the one-band Hubbard 
model requires $U >W$.

Calculations within the Dynamical Mean Field Approximation 
(DMFA) \cite{metzvoll} can be performed in the strong coupling regime
and take place in the thermodynamic limit. But the lack of non-local correlations inhibits a
transition to a state with a non-local ($d$-wave) order parameter. The
recently developed Dynamical Cluster Approximation (DCA)
\cite{DCA_hettler,DCA_maier1,huscroft} is a fully causal approach which
systematically incorporates non-local corrections to the DMFA by
mapping the lattice problem onto an embedded periodic cluster of size $N_c$.
For $N_c=1$ the DCA is equivalent to the DMFA and by increasing the
cluster size $N_c$ the dynamic correlation length can be gradually
increased while the DCA solution remains in the thermodynamic limit. 

Using a Nambu-Gorkov representation of the DCA we observe a transition
to a superconducting phase in doped systems at sufficiently low
temperatures. This 
occurs in the intermediate to strong coupling regime $U>W$ and the corresponding order parameter has $d$-wave symmetry.

\paragraph*{Method}

\begin{figure}[ht]
\leavevmode\centering\psfig{file=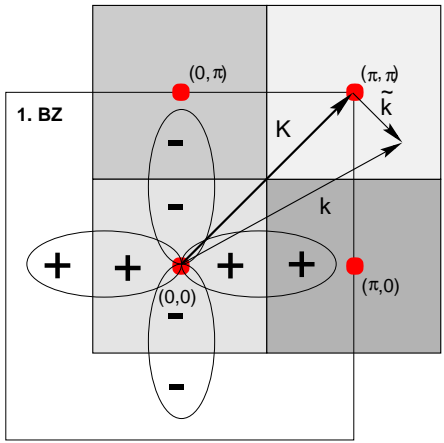,width=2.3 in}

\vspace{0.5cm}\caption{Choice of the $N_c=4$ cluster $\K$-points
(filled circles), corresponding coarse graining cells (shown by
different fill patterns) and a sketch of the $d$-wave symmetry of the
order parameter.} 
\label{BZsc}       
\end{figure}

A detailed discussion of the DCA formalism was given in previous 
publications\cite{DCA_hettler,DCA_maier1,huscroft} where it was shown to 
systematically restore momentum conservation at internal diagrammatic
vertices which is relinquished by the DMFA.  However, the DCA also has
a simple physical interpretation based on the observation that the self 
energy is only weakly momentum dependent for systems where the dynamical 
intersite correlations have only short spatial range.  The corresponding 
self-energy is a functional of the interaction $U$ and the Green function 
propagators.  The latter may be calculated on a coarse grid of $N_c=L^D$ 
selected $\K$-points only, where $L$ is the 
linear dimension of the cluster of $\K$-points. According to Nyquist's 
sampling theorem \cite{nyquist}, this sampling of the reciprocal space at 
intervals of $\Delta\k=2\pi/L$ implies that the DCA incorporates nonlocal 
dynamical correlations with a spatial range $\alt L/2$ and cuts off 
longer ranged dynamical correlations. Knowledge of the momentum dependence 
on a finer grid may be discarded to reduce the complexity of the problem.  
To this end the first Brillouin zone is divided into $N_c$ cells of size 
$(2\pi/L)^D$ around the cluster momenta $\K$ (see Fig.\ref{BZsc}). The Green 
functions used to form the self-energy $\Sigma(\K,\omega)$ are coarse 
grained, or averaged over the momenta $\K+\tk$ surrounding the cluster 
momentum points $\K$ (cf. Fig.\ref{BZsc}).

Thus, the {\em coarse grained} Green function is
\be
\hat{\bar{G}}(\K,\omega)=\frac{N_c}{N}\sum_{\tk}\hat{G}(\K+\tk,\omega)\quad\mbox{,}
\label{cgG}
\ee
where the sum runs over all vectors $\k=\K+\tk$ within a cell around
the cluster momentum $\K$.  Note that the choice of the coarse grained 
Green function has two well defined limits: For $N_c=1$ the sum over $\tk$ 
runs over the entire Brillouin zone, $\hat{\bar{G}}$ is the local Green 
function, thus the DMFA algorithm is recovered. For $N_c=\infty$ the 
$\tk$-summation vanishes and the DCA becomes equivalent to the exact 
solution.  The dressed lattice Green function takes the form
\be
\hat{G}(\k,\omega)=\left(\omega {\mathbbm
1}-\ek\tau_3-\hat{\Sigma}(\K,\omega)\right)^{-1}\quad\mbox{,}
\label{NGL}
\ee
with the self-energy $\hat{\Sigma}(\k,\omega)$ approximated by the cluster
self-energy $\hat{\Sigma}(\K,\omega)$. To allow for a possible transition 
to the superconducting state we utilized the Nambu-Gorkov matrix
representation \cite{Schrieffer} in (\ref{NGL}) where the self-energy matrix
$\hat{\Sigma}$ is most generally written as an expansion 
$\hat{\Sigma}=\sum_i \Sigma_i \tau_i$ in terms of the Pauli matrices 
$\tau_i$. The diagonal components of $\hat{\Sigma}$ represent quasiparticle 
renormalizations, whereas the offdiagonal parts are nonzero in the
superconducting state only.

Since the self-energy $\hat{\Sigma}(\K,\omega)$ does not depend on the 
integration variable $\tk$, we can write 
\be
\hat{\bar{G}}(\K,\omega)=(\omega {\mathbbm
1}-\eK\tau_3-\hat{\Sigma}(\K,\omega)-\hat{\Gamma}(\K,\omega))^{-1}\mbox{,}
\label{cgNCA}
\ee 
where $\eK=N_c/N \sum_{\tk} \epsilon_{\K+\tk}$. 
This has the form of the Green function of a cluster model with
periodic boundary conditions coupled to a dynamic host described by
$\hat{\Gamma}(\K,\omega)$. Here we employ
the NCA to calculate the cluster Green function and self-energy
respectively. A detailed discussion of the NCA-algorithm applied to the cluster
model for the paramagnetic state was given in \cite{DCA_maier1}. 
The NCA for the superconducting state has to be
extended in order to account for the hybridization to the anomalous
host, which couples cluster states with different particle
numbers.

The self-consistent iteration is initialized by calculating the
coarse grained average $\hat{\bar{G}}(\K)$ (Eq. \ref{cgG}) and with
Eq. \ref{cgNCA} the host function $\hat{\Gamma}(\K)$, which is used
as input for the NCA. The NCA result for the cluster
self-energy $\hat{\Sigma}(\K)$ is then used to calculate a new
estimate for the coarse grained average $\hat{\bar{G}}(\K)$
(Eq. \ref{cgG}).  The procedure
continues until the self-energy converges to the desired accuracy.

\paragraph*{Results}
We investigate the single particle properties of the doped 2D Hubbard Model
\be
H=\sum_{ij,\sigma}t_{ij}c_{i\sigma}^\dagger c^{}_{j\sigma}+U\sum_i n_{i\uparrow}n_{i\downarrow}\quad\mbox{,}
\ee
where $c^\dagger_{i}$ ($c^{}_{i}$) creates (destroys) an electron at
site $i$ with spin $\sigma$ and $U$ is the on-site Coulomb
repulsion. For the Fourier transform of the hopping
integral $t_{ij}$ we use
\be
\ek=\epsilon_o-\mu-2t(\cos k_x+\cos k_y)-4t^\prime \cos k_x \cos k_y\mbox{,}
\ee 
accounting for both, nearest neighbor hopping $t$ and next
nearest neighbor hopping $t^\prime$. We set $t=0.25{\rm eV}$ and
$U=3{\rm eV}$, well above the
bandwidth $W=8t=2{\rm eV}$. For this choice of parameters the system
is a Mott-Hubbard insulator at half filling as required for a proper
description of the high-$T_c$ cuprates.

To allow for symmetry breaking we start the iteration procedure with
finite offdiagonal parts of the self-energy matrix $\hat{\Sigma}$. As we mentioned above one expects the order parameter of a possible
superconducting phase to have $d$-wave symmetry. 
Therefore we work with a 2x2-cluster ($N_c=4$), the
smallest cluster size incorporating nearest neighbor correlations. For
the set of cluster points we choose $\K_{\alpha l}=l\pi$, where
$l=0,1$ and $\alpha=x$ or $y$. 
Fig.\ref{BZsc} illustrates this choice of $\K$-points along with a sketch of
the $d$-wave order parameter and the coarse graining
cells. Obviously, for symmetry reasons, in the case of $d$-wave superconductivity, we expect the coarse grained
anomalous Green function to vanish at the zone center and the point
$(\pi,\pi)$. Whereas the anomalous parts at the points $(0,\pi)$ and
$(\pi,0)$ should be finite with opposite signs.    
\begin{figure}[ht]
\leavevmode\centering\psfig{file=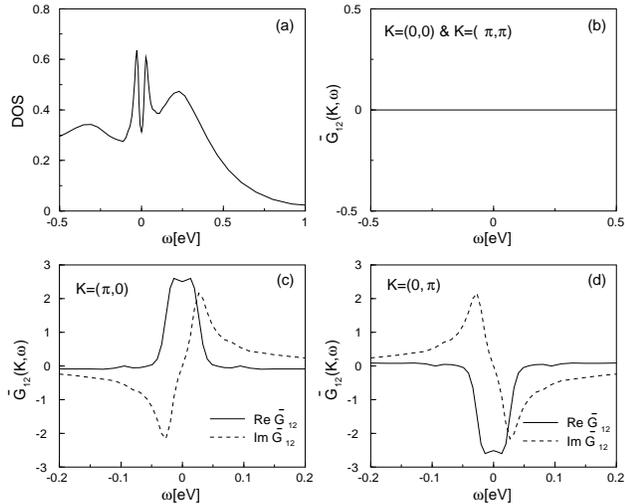,width=3.25 in}

\vspace{0.5cm}
\caption{(a) The local density of states (DOS) near the Fermi energy and the anomalous coarse grained
Green functions at the cluster points (b) $\K=(0,0)$ and $\K=(\pi,\pi)$, (c)
$\K=(\pi,0)$ and (d) $\K=(0,\pi)$ in the superconducting state. The nearest neighbor hopping integral
$t=0.25{\rm eV}$, next nearest neighbor hopping integral $t^\prime=0$,
bandwidth $W=2{\rm eV}$, the on-site Coulomb repulsion $U=3{eV}$,
temperature $T=137{\rm K}$ and
the doping $\delta=0.19$. The anomalous parts of the Green function
(b)-(d) are consistent with a $d$-wave order parameter.}  
\label{res1}       
\end{figure}
Fig.\ref{res1} shows a typical result for the local density of states (DOS)
in the superconducting state along with the anomalous coarse
grained Green function $\bar{G}_{12}(\K,\omega)=N_c/N \sum_{\tk}
\langle\langle c_{\K+\tk\uparrow};c_{-(\K+\tk)\downarrow}\rangle\rangle_\omega$ at the cluster $\K$-points for $t^\prime =0$,
temperature $T=137{\rm K}$ and doping $\delta=0.19$.  
The anomalous coarse grained Green function vanishes at the cluster points
$(0,0)$ and $(\pi,\pi)$ but is finite at the points $(\pi,0)$
and $(0,\pi)$, consistent with
a $d$-wave order parameter. Note that this result is independent of the initialization of the
self-energy, i.e. an additional initial $s$-wave contribution vanishes 
in the course of the iteration. Thus a possible $s$-wave contribution to the order
parameter can be ruled out.

The finite pair
amplitude is also reflected in the local density of states (DOS) depicted in
Fig.\ref{res1}a, where we show the lower sub-band of the full spectrum
near the Fermi energy. It displays a pseudogap at
zero frequency as expected for a $d$-wave order parameter.
                                                            
\begin{figure}[ht]
\leavevmode\centering\psfig{file=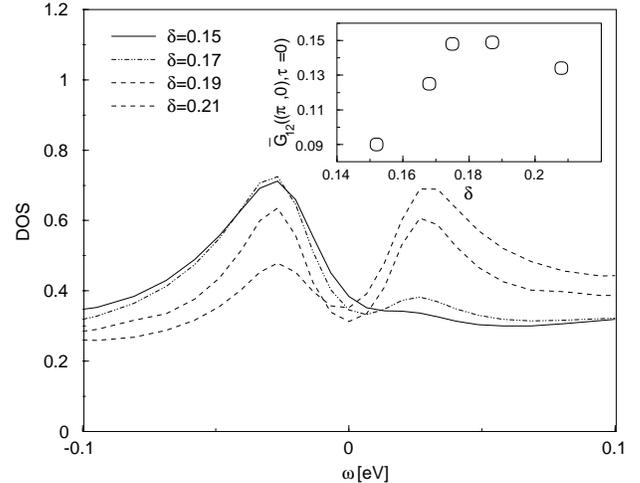,width=3.2 in}

\vspace{0.5cm}\caption{Density of states in a narrow region
at the Fermi energy for the same parameters as in Fig.\ref{res1} but
for various dopings. The gap size and the density of states at
$\omega=0$ are independent of doping. Inset: Equal time coarse grained 
anomalous Green function $\bar{G}_{12}(\K,\tau=0)$ at
$\K=(\pi,0)$.}  
\label{res2}       
\end{figure}

Fig.\ref{res2} shows the DOS near the Fermi energy for the same
parameters as in Fig.\ref{res1}, fixed temperature $T=137{\rm K}$, but for
various dopings. Obviously, the pseudogap size, measured as the peak
to peak distance, as well as the density of states at the Fermi 
energy do not depend strongly upon doping. However the drop in the
density of states from the gap edge to the $\omega=0$ value first
increases, reaches a maximum at about 19\% doping, then decreases
again.
This behavior originates in the doping dependence of the anomalous Green
function. In the inset we plot the coarse grained anomalous equal time 
Green function
$\bar{G}_{12}(\K,\tau=0)=N_c/N\sum_{\tk}\langle c_{\K+\tk\uparrow}
c_{-(\K+\tk)\downarrow}\rangle$ for $\K=(\pi,0)$. This number as a
measure of the superconducting gap shows exactly the same behavior as
the pseudogap in the density of states.   

\begin{figure}[ht]
\leavevmode\centering\psfig{file=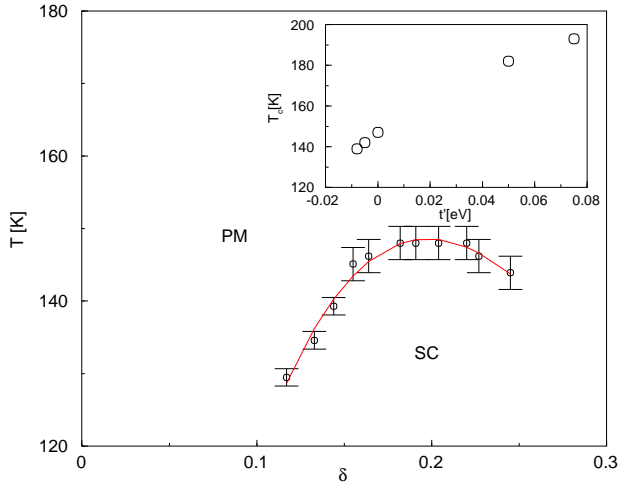,width=3.2 in}

\vspace{0.5cm}\caption{Temperature-doping phase diagram for the 2D
Hubbard model via DCA for a $N_c=4$ cluster. The nearest neighbor
hopping $t=0.25{\rm eV}$, next nearest neighbor hopping $t^\prime=0$
and the Coulomb repulsion $U=3{\rm eV}$. The error bars result from
the finite resolution in temperature. Inset: Transition temperature 
$T_c(t^\prime)$ for fixed doping $\delta=0.18$ as a function of the
next nearest neighbor hopping amplitude $t^\prime$.}  
\label{res3}       
\end{figure}

The anomalous components $\bar{G}_{12}(\K,\omega)$ and hence the
pseudogap in the DOS become smaller with increasing temperature and
eventually vanish at a critical temperature $T_c$ depending on the set 
of parameters.
The phase diagram is shown in Fig.\ref{res3}.
As a function of doping, $T_c(\delta)$ has a maximum $T_c^{max}\approx 
150{\rm K}$ at $\delta\approx 19\%$ and strongly decreases with
decreasing or increasing $\delta$. The qualitative behavior of
$T_c(\delta)$ in the calculated $T-\delta$ region agrees well with the 
generic phase diagram of the high-$T_c$ cuprates. Unfortunately, due
to the break-down of the NCA at very low temperatures we are not able
to extend the phase diagram beyond the region shown in
Fig.\ref{res3}. This means in particular that we cannot predict reliable
values for $\delta_c(T=0)$, beyond which superconductivity vanishes.

\begin{figure}[ht]
\leavevmode\centering\psfig{file=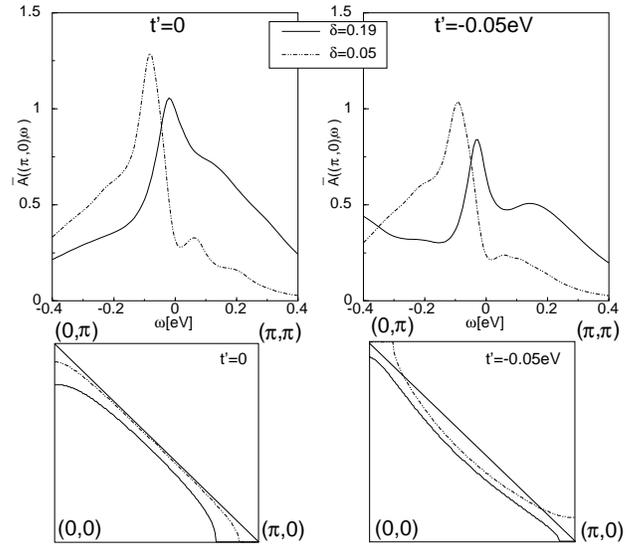,width=3.2 in}

\vspace{0.5cm}
\caption{Coarse grained spectral density  
$-\frac{1}{\pi} \Im m
\bar{G}_{11}(\K,\omega)$ at $\K=(\pi,0)$ in the normal state
($T=290{\rm K}$) for the next nearest
neighbor hopping $t^\prime=0$ (left) and $t^\prime=-0.05eV$
(right) for different dopings $\delta$. The nearest neighbor hopping $t=0.25eV$ and the Coulomb
repulsion $U=3eV$. Bottom: Fermi surfaces of the corresponding
noninteracting ($U=0$) systems in the first quadrant of the BZ. The
thick diagonal line indicates the set of $\k$-points which
fulfill the nesting condition.
}  
\label{res4}       
\end{figure}

The inset of Fig.\ref{res3} shows the transition temperature
dependence $T_c(t^\prime,\delta={\rm const.})$  on the next nearest neighbor hopping amplitude $t^\prime$
for fixed doping $\delta=0.18$. As compared to $t^\prime=0$ $T_c$
strongly decreases with growing negative $t^\prime$ but increases for $t^\prime>0$. The shape of the phase diagram as well as the $t^\prime$-dependence of 
$T_c$ can be qualitatively understood in terms of the phenomenological 
picture, where spin
fluctuations mediate the electron-electron interaction, which then is strong at the
antiferromagnetic wave vector ${\bf Q}=(\pi,\pi)$. 
In Fig.\ref{res4} we display the coarse grained spectra
$-\frac{1}{\pi}\Im m\bar{G}_{11}(\K,\omega)$ at $\K=(\pi,0)$ in the 
normal state ($T=290{\rm K}$) for the next nearest
neighbor hopping $t^\prime=0$ and $t^\prime=-0.05eV$
(left and right hand side) for different dopings $\delta$.
At the
bottom we show 
the Fermi surfaces of the corresponding noninteracting systems
($U=0$) in the first quadrant of the BZ. The diagonal thick solid line indicates the set of
$\k$-points which fulfill the nesting condition, i.e. which can be
connected by $\bf{Q}$ to equivalent
$\k$-points in the opposite quadrant of the BZ. 

The doping dependence
of the $t^\prime=0$ and $t^\prime=-0.05$ spectra
is qualitatively different. In the $t^\prime=-0.05$ case the parts of 
the Fermi surface near $\K=(\pi,0)$ and $(0,\pi)$ fulfill the
nesting condition roughly for the whole doping range, the
quasiparticles couple strongly to the spin fluctuations and hence the
corresponding spectra display a
pseudogap at zero frequency over the entire doping range. The
$t^\prime=0$ spectra in contrast exhibit the pseudogap in the underdoped regime
only ($\delta=0.05$), where the spin fluctuations are strong, but show a quasiparticle
peak at optimal doping $\delta=0.19$, where the points on the Fermi
surface near $\K=(\pi,0)$ and $(0,\pi)$ are far from being nested. The suppression of the density of states at the Fermi energy
results in a suppression of superconductivity and hence the transition
temperatures drop with decreasing doping as well as decreasing
$t^\prime<0$. For positive $t^\prime$ we obtain similar spectra and
Fermi surfaces as for $t^\prime=0$, but with a slightly enhanced density of
states at the Fermi energy, resulting in higher transition temperatures. 
        
\paragraph*{Summary} We have used the recently developed DCA to study 
the long open question of whether the  2D Hubbard model shows
instabilities towards a superconducting
state in the intermediate to strong coupling regime. We find
conclusive evidence that at moderate doping a transition to a state
with offdiagonal long range order occurs and that the corresponding
order parameter has pure $d$-wave symmetry. The corresponding
temperature-doping phase diagram agrees qualitatively with the generic
high-$T_c$ phase diagram.

\paragraph*{Acknowledgements} 
It is a pleasure to acknowledge useful discussions with P.G.J.~van
Dongen, M.~Hettler and H.R.~Krishnamurthy. This work was supported by NSF grants DMR-9704021, DMR-9357199 and the 
Graduiertenkolleg ``Komplexit\"at in Festk\"orpern''. Computer support 
was provided by the Ohio Supercomputer Center and the
Leibnitz-Rechenzentrum, Munich.


\begin{references}

\bibitem{maple1} For a review, see M.B.~Maple, cond-mat/980202.

\bibitem{ding} H.~Ding {\it et al.,} Nature {\bf 382}, 51 (1996).

\bibitem{ronning} F.~Ronning {\it et al.,} Science (1998).

\bibitem{scalapino1} For a review, see D.J.~Scalapino, cond-mat/9908287.

\bibitem{Monthoux} P.~Monthoux, A.V.~Balatsky and D. Pines,
Phys. Rev. Lett., {\bf 67}, 3448, (1991).

\bibitem{timusk} T.~Timusk and B.~Statt, cond-mat/9905219.

\bibitem{Anderson} P.W.~Anderson, {\bf The Theory of Superconductivity 
in the High-$T_c$ Cuprates}, Princeton University Press , Princeton,
NJ (1997)

\bibitem{bickers} N.E.~Bickers, D.J.~Scalapino, and S.R.~White,
Phys. Rev. Lett., {\bf 62}, 961 (1989).

\bibitem{Moriya} T.~Moriya, Y.~Takahashi, and K.~Ueda,
J. Phys. Soc. Japan, {\bf 59}, 2905, (1990).

\bibitem{metzvoll} W.~Metzner and D.~Vollhardt, Phys.\ Rev.\ Lett.\
{\bf 62}, 324 (1989).

\bibitem{DCA_hettler} M.H.~Hettler {\it et al.}, Phys.\ Rev.\ B {\bf  58}, 7475 
(1998).   M.H.~Hettler {\it et al.,} preprint cond-mat/9903273.

\bibitem{DCA_maier1} Th.~Maier {\it et al.},
Eur.\ Phys.\ J.\ B {\bf 13}, 613 (2000).

\bibitem{huscroft} C.~Huscroft et al., cond-mat/9910226.

\bibitem{nyquist} D.F.~Elliot and K.R.~Rao, {\em{Fast Transforms:
Algorithms, Analyses, Applications}} (Academic Press, New York, 1982).



\bibitem{Schrieffer} J.R.~Schrieffer, {\bf Theory of
Superconductivity}, Addison Wesley, Reading, MA (1993).

\bibitem{keiter} H.~Keiter, J.C.~Kimball, Intern. J. Magnetism, {\bf 1}, 233 (1971).




\end{references}
\end{document}